\documentclass[proceedings, preprint]{rmaa}

\usepackage{paralist}

\usepackage{psfrag,color}

\SetYear{2022}
\SetConfTitle{Prospects for low-frequency radio astronomy in S. A.}

\title{Pulsar Observations at low latitudes and low frequencies} 

\author{Carlos O. Lousto,\altaffilmark{1} \\
  R.Missel,\altaffilmark{2}
  E.Zubieta,\altaffilmark{3,4} 
  S.del Palacio,\altaffilmark{4,5}
  F.Garcia,\altaffilmark{4}
  G.Gancio,\altaffilmark{4}
  L.Wang,\altaffilmark{2}
  S.B.Araujo Furlan,\altaffilmark{6,7}
  J.A.Combi\altaffilmark{8} 
}

\altaffiltext{1}{Center for Computational Relativity and Gravitation (CCRG), School of Mathematical Sciences, Rochester Institute of Technology, 85 Lomb Memorial Drive, Rochester, New York 14623, USA.}
\altaffiltext{2}{Golisano College of Computing and Information Sciences, Rochester Institute of Technology Rochester, NY 14623, USA.}
\altaffiltext{3}{Facultad de Ciencias Astron\'omicas y Geof\'{\i}sicas, Universidad Nacional de La Plata, Paseo del Bosque, B1900FWA La Plata, Argentina.}
\altaffiltext{4}{Instituto Argentino de Radioastronom\'ia (CCT La Plata, CONICET; CICPBA; UNLP), C.C.5, (1894) Villa Elisa, Buenos Aires, Argentina.}
\altaffiltext{5}{Department of Space, Earth and Environment, Chalmers University of Technology, Sweden. SE-412 96 Gothenburg.}
\altaffiltext{6}{Instituto de Astronom\'\i{}a Te\'\o{}rica y Experimental, CONICET-UNC, Laprida 854, X5000BGR – C\'ordoba, Argentina.}
\altaffiltext{7}{Facultad de Matem\'atica, Astronom\'{i}a, F\'{i}sica y Computaci\'on, UNC. Av. Medina Allende s/n, Ciudad Universitaria, CP:X5000HUA - C'ordoba, Argentina. }
\altaffiltext{8}{Departamento de F\'\i sica (EPS), Universidad de Ja\'en, Campus Las Lagunillas s/n, A3, 23071 Ja\'en, Spain.}

\shortauthor{Lousto, PuMA Collaboration} 
\shorttitle{Pulsar Observations in South America}

\listofauthors{Lousto, PuMA Collaboration}
\indexauthor{Lousto, PuMA Collaboration}

\abstract{The Pulsar Monitoring in Argentina (PuMA) is a collaboration between the Argentine Institute for Radioastronomy (IAR) and the Rochester Institute of Technology (RIT) that since 2017 has been observing southern sky pulsars with high cadence using the two restored IAR antennas in the L-Band (1400MHz).
We briefly review the first set of results of this program to study transient phenomena, such as magnetars and glitching pulsars, as well as to perform precise timing of millisecond pulsars. Access to lower frequency bands, where most of the pulsars are brighter, would allow us to reach additional pulsars, currently buried into the background noise. We identify two dozen additional glitching pulsars that could be observable in the 400MHz band by the  IAR's projected Multipurpose Interferometer Array (MIA). We also discuss the relevance and challenges of single-pulse pulsar timing at low frequencies and the search for Fast Radio Burst (FRB) in the collected data since 2017 using machine learning techniques.}

\resumen{La colaboraci\'on PuMA (Pulsar monitoring in Argentina) entre el Instituto Argentino de Radioastronom\'{i}a (IAR) y el Rochester Institute of Technology (RIT) ha estado observando p\'ulsares en el hemisferio sur desde el a\~no 2017 con una cadencia aproximadamente diaria utilizando las dos antenas recientemente restauradas del IAR en la banda-L (1400MHz). Aqu\'{i} presentamos una breve rese\~na de los primeros resultados del programa PuMA para observar fen\'omenos transitorios, como los magnetares y los p\'ulsares con anomal\'{i}as en sus per\'{i}odos, as\'{i} como medidas muy precisas del tiempo de llegada de los pulsos provenientes de p\'ulsares con per\'{i}odos de milisegundos. El accesso a observaciones en m\'as baja radiofrecuencias, donde la major\'{i}a de los p\'ulsares tienen un espectro de radiaci\'on m\'as brillante, nos permitir\'{i}a observar con suficiente precisi\'on nuevos p\'ulsares, que actualmente presentan demasiado ruido de fondo en la banda-L. As\'{i}, identificamos una docena de p\'ulsares de inter\'es, que presentan anomal\'{i}as en sus per\'{i}odos, y que podr\'{i}an ser observado por el nuevo instrumento proyectado por el IAR, el Multipurpose Interferometer Array (MIA), en la banda de los 400MHz. Tambi\'en discutimos aqu\'{i} la importancia de las observaciones y el estudio de pulsos individuales (y sus dificultades) para mejorar la precisi\'on de la medida de los tiempos de arrivo de los pulsos de p\'ulsares de milisegundos y la aplicaci\'on de t\'ecnicas de aprendizaje autom\'atico de inteligencia artificial para la b\'usqueda de FRBs (Fast Radio Burst -- r\'afagas r\'apidas de radio) en la gran cantidad de datos colectados por el IAR desde el a\~no 2017.}

\addkeyword{Instrumentation: detectors}
\addkeyword{methods: observational}
\addkeyword{methods: statistical}
\addkeyword{pulsars: Vela}
\addkeyword{pulsars: PSR J0437$-$4715}

\begin{document}

\maketitle

\section{Introduction}\label{sec:intro}

\begin{figure}[ht]
  \centering
    \includegraphics[width=\linewidth]{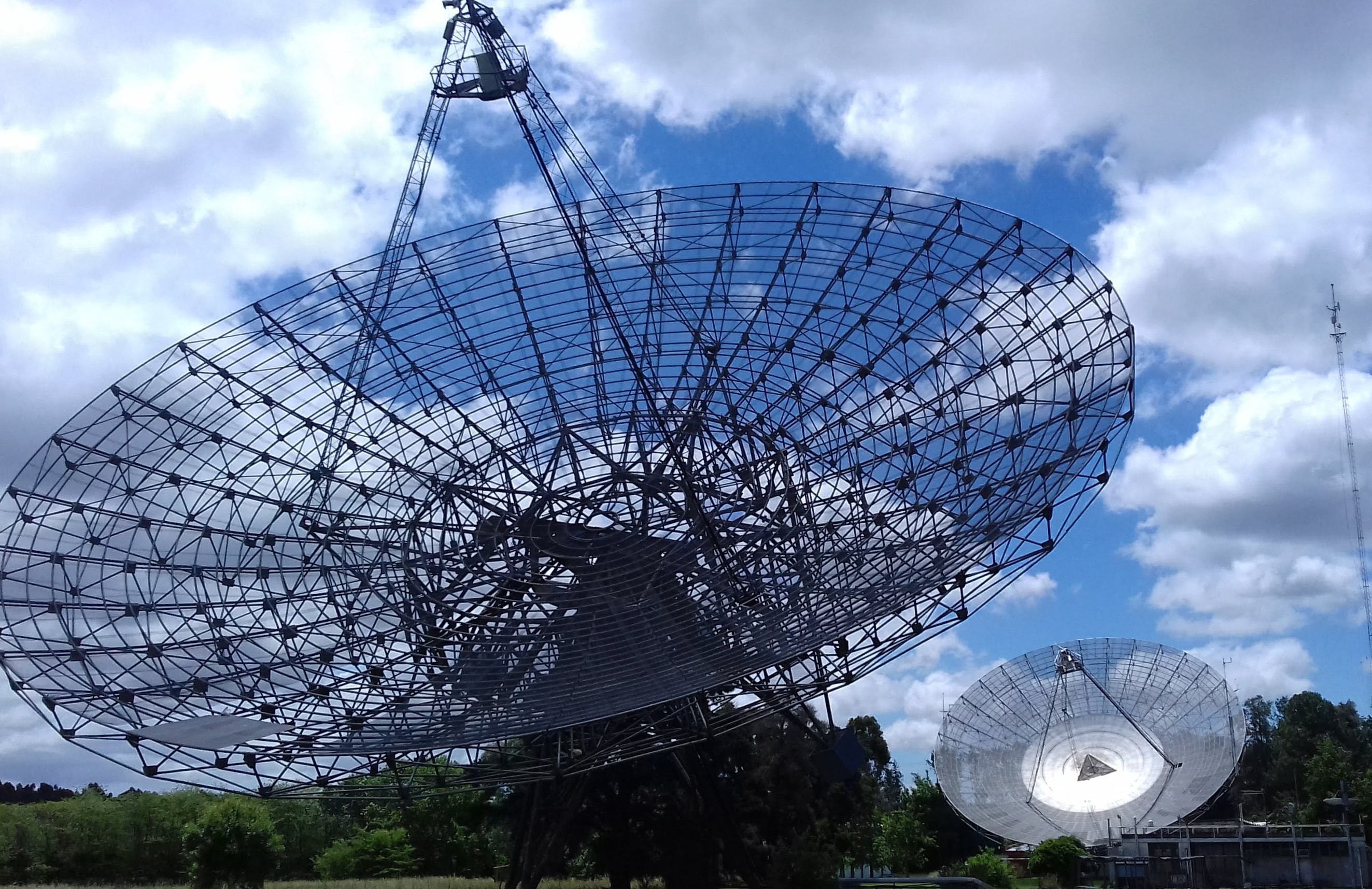}
    \caption{View of IAR antennas, A2 (left) and A1 (right).}
\label{fig:AntennaII&I}
\end{figure}

The Argentine Institute of Radio astronomy (IAR) was founded in 1962 as a pioneer radio observatory in South America with two 30-meter parabolic single-dish radio antennas (Fig.~\ref{fig:AntennaII&I}). Antenna 1 (A1) saw its first light in 1966 whereas Antenna 2 (A2) was built later in 1977. The IAR's initial purpose was to perform a high sensitivity survey of neutral hydrogen ($\lambda=21$~cm) in the southern hemisphere; this survey ended satisfactorily in the year 2000 with high-impact publications in collaboration with German and Dutch institutions \citep{Testori:2001vp,Bajaja:2005tn,Kalberla:2005ts}.
(See G. Romero's contribution to this volume for more historical details).

Although the IAR has been a center of intense scientific and technological activity since its foundation, the radio antennas had not been employed in any scientific project since 2001. 
The PuMA\footnote{\url{http://puma.iar.unlp.edu.ar}} (Pulsar Monitoring in Argentina) is a new collaboration of scientists and technicians from the IAR and the Rochester Institute of Technology (RIT) dedicated to the formation of human resources for observations, data analysis, and pulsar astrophysics studies. This project represents the first systematic pulsar timing observations in South America and the beginning of pulsar science in Argentina. 
(See P. Benaglia's contribution to this volume for other independent research projects being carried out at IAR).

Since 2017 the IAR antennas have been upgraded to conduct high-quality radio astronomy \citep{Gancio:2019frj}
to enable science projects for the first time in over fifteen years. These projects include: i) Pulsar timing and gravitational waves, ii) Targeted pulsar studies for continuous gravitational waves detection from laser interferometry,
iii) Magnetars, iv) Glitches and young pulsars, v) Fast-radio-burst observations, vi) Interstellar medium scintillation, vii) Tests of gravity with pulsar timing. 
%


\section{Main Results}\label{sec:Results}

\subsection{J0437$-$4715 first timing campaign}

We presented the first-year of high-cadence, long-duration observations of the bright millisecond pulsar J0437$-$4715 obtained in the IAR in \citep{SosaFiscella:2020wmm}. Using the two single-dish 30 m radio antennas, we gathered more than 700 hr of good-quality data with timing precision better than 1~$\mu$s. We characterized the white and red timing noise in IAR's observations, we quantified the effects of scintillation, and we performed single-pulsar searches of continuous gravitational waves, setting constraints in the nHz--$\mu$Hz frequency range. We thus demonstrated IAR's potential for performing pulsar monitoring in the 1.4 GHz radio band for long periods of time with a nearly daily cadence. In particular, we concluded that the ongoing observational campaign of J0437$-$4715 can contribute to increase the sensitivity of the existing pulsar-timing arrays.

The characterization of the observations used in this work are sumarized in Table \ref{table:par_obs}.
\begin{table}[htb]
	{\centering
	\caption{J0437$-$4715 Observations} \label{table:par_obs}
	\begin{tabular}{lcc}
		\hline \hline
		 & A1 & A2 \\
		\hline
        Number of observations & 170 & 197 \\
		MJD start--MJD finish & \multicolumn{2}{c}{58596.7 -- 58999.6} \\
		Total observation time [h] & 391 & 393 \\
		Central frequency [MHz] & 1400-1428 & 1428  \\
		Bandwidth ($BW$) [MHz] & 112 & 56 \\
		Polarization modes & 1 & 2  \\
		Frequency channels ($n_\mathrm{chan}$) & 64/128 & 64 \\
		Time resolution  [$\mu$s] & {73.14} & {73.14}\\
		Phase bins ($n_\mathrm{bin}$) &512/1024&512/1024\\
		\hline \hline
	\end{tabular}
	\par}
\end{table}

\begin{figure}[htb]
    \centering
    \includegraphics[width=\linewidth]{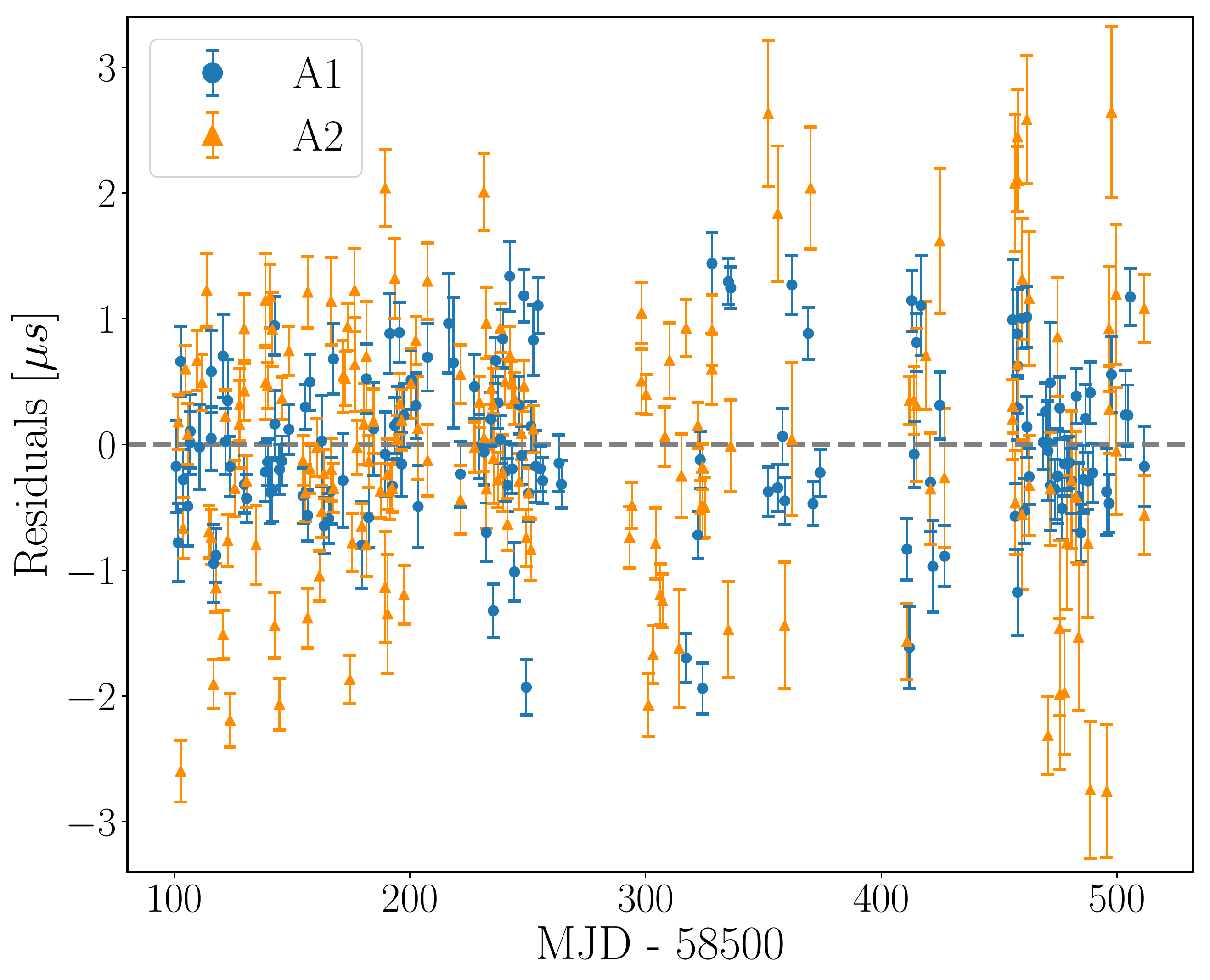}
    \caption{Timing residuals for the complete data set for A1 and A2.}
    \label{fig:residues_total}
\end{figure}

Fig.~\ref{fig:residues_total} shows the timing residuals of the observations taken with each antenna.

\begin{figure}[htb]
    \centering
    \includegraphics[width=\linewidth]{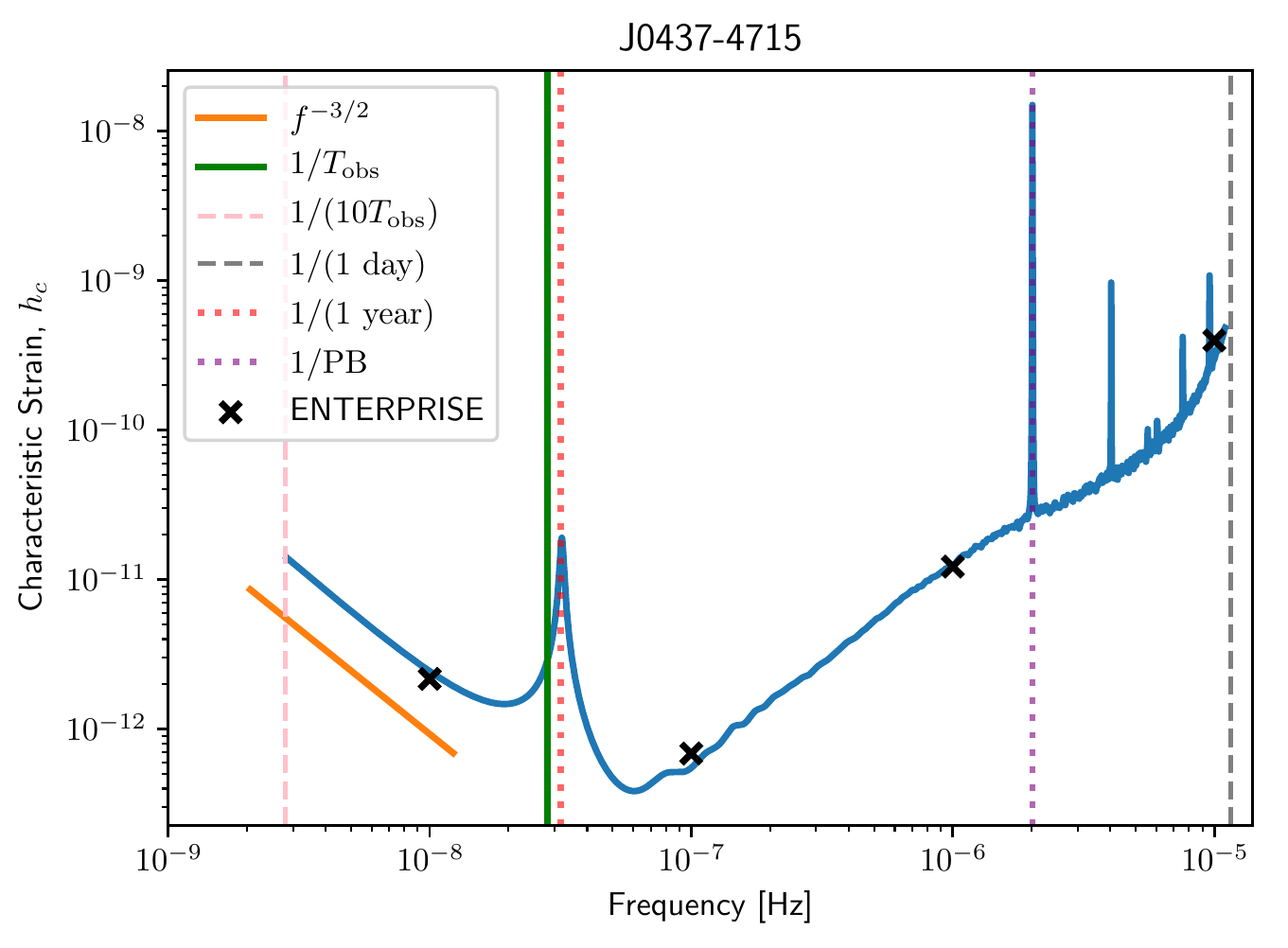}
    \caption{ 
    Sensitivity curve for J0437$-$4715 using 1.1~yr observations at IAR, including pulsar noise characteristics, for a single deterministic gravitational wave source averaged over its initial phase, inclination, and sky location (A1+A2; blue curve). The vertical green line corresponds to a frequency of $1/T_{\mathrm{obs}}$, the dotted red line to $1/T_{\mathrm{yr}}$ and the dotted purple line to $1/{\mathrm{PB}}$ (orbital period). The black crosses correspond to the mean values of the $\log_{10} h_\mathrm{gw}$ distributions obtained using \texttt{ENTERPRISE}.}
    \label{fig:sensitivity_J04}
\end{figure}

A single supermassive binary black-hole system produces “continuous” gravitational wavess because the system does not evolve notably over the few years of a pulsar-timing data set. We used the Python package \texttt{Hasasia} \citep{Hazboun2019Hasasia} to calculate the single-pulsar sensitivity curve of our data set of J0437$-$4715 for detecting a deterministic gravitational waves source averaged over its initial phase, inclination, and sky location. 
The resulting sensitivity curve is shown in Fig.~\ref{fig:sensitivity_J04}. It is readily seen that there is a loss of sensitivity at a frequency of $(1~\mathrm{yr})^{-1}$, caused by fitting the pulsar's position, and at a frequency of $(\mathrm{PB})^{-1} \sim 2~\mathrm{\mu}$Hz (with $\mathrm{PB}$ the orbital period), caused by fitting the orbital parameters of the binary system. The additional spikes seen at frequencies higher than $(\mathrm{PB})^{-1}$ correspond to harmonics of the binary orbital frequency.

For comparison, we used the code \texttt{ENTERPRISE} \citep{Ellis2019} to perform a fixed-frequency Markov chain Monte Carlo procedure at four different frequencies. We obtained a posterior distribution for $\log_{10} h_\mathrm{gw}$ at each of these frequencies with a mean value in great agreement with the curve obtained with \texttt{Hasasia}, as shown in Fig.~\ref{fig:sensitivity_J04}. 

These first results on gravitational waves sensitivity are encouraging, though we still need to achieve a sensitivity of at least a factor 10 higher in order to observe even the most favourable supermassive black hole binary merger events. For instance, the six billion solar mass source of 3C~186 at $z\approx1$ produced a gravitational wave amplitude of $h\sim 10^{-14}$ at the time of arrival to our Galaxy, roughly a million years ago~\citep{Lousto:2017uav}.
The goal of the PuMA collaboration is to continue analyzing the additional years since 2020 of J0437$-$4715 observations using the traditional timing techniques as well as the single pulse studies, as performed for Vela pulsar \citep{Lousto:2021dia}.


\subsection{Single pulses analysis with machine learning techniques: Vela pulsar}

In \citep{Lousto:2021dia} we studied individual pulses of Vela (PSR\ B0833-45\,/\,J0835-4510) from daily observations of over three hours (around 120,000 pulses per observation), performed simultaneously with the two radio telescopes at the IAR. We selected 4 days of observations in January-March 2021 and study their statistical properties with machine learning techniques. We first used density based DBSCAN clustering techniques, associating pulses mainly by amplitudes, and find a correlation between higher amplitudes and earlier arrival times. We also found a weaker correlation with the mean width of the pulses. We identified clusters of the so-called mini-giant pulses, with $\sim10$ times the average pulse amplitude. We then performed an independent study, with Self-Organizing Maps (SOM) clustering techniques. We use Variational AutoEncoder (VAE) reconstruction of the pulses to separate them clearly from the noise and select one of the days of observation to train VAE and applied it to the rest of the observations. We used SOM to determine 4 clusters of pulses per day per radio telescope and concluded that our main results are robust and self-consistent. These results support models for emitting regions at different heights (separated each by roughly a hundred km each) in the pulsar magnetosphere.
We also modeled the pulses amplitude distribution with interstellar scintillation patterns at the inter-pulses time-scale finding a characterizing exponent $n_{\mathrm{ISS}}\sim7-10$.

We analyzed observations on January, 21th, 24th and 28th, and on March 29th, 2021, performed concurrently with both radio telescopes for over three hours. The number of single-pulses in each observation is given in Table~\ref{tab:observations}.
\begin{table}
	\centering
	\caption{Date, number of single pulses, MJD and instantaneous period at the beginning of each observation}
	\label{tab:observations}
	\begin{tabular}{llcl} 
		\hline
		Day 2021&initial MJD&\#pulses&$P_{\mathrm{inst}}$ [ms]\\
		\hline
		Jan. 21 & 59235.128553 & 121495 & 89.407366\\
		Jan. 24 & 59239.117013 & 119448 & 89.407676\\
		Jan. 28 & 59242.088680 & 128999 & 89.407915\\
		Mar. 29 & 59302.943356 & 128999 & 89.413017\\
		\hline
	\end{tabular}
\end{table}

\subsubsection{Scintillation}\label{sec:scintillations}
\begin{figure}[htb]
	\includegraphics[width=\columnwidth]{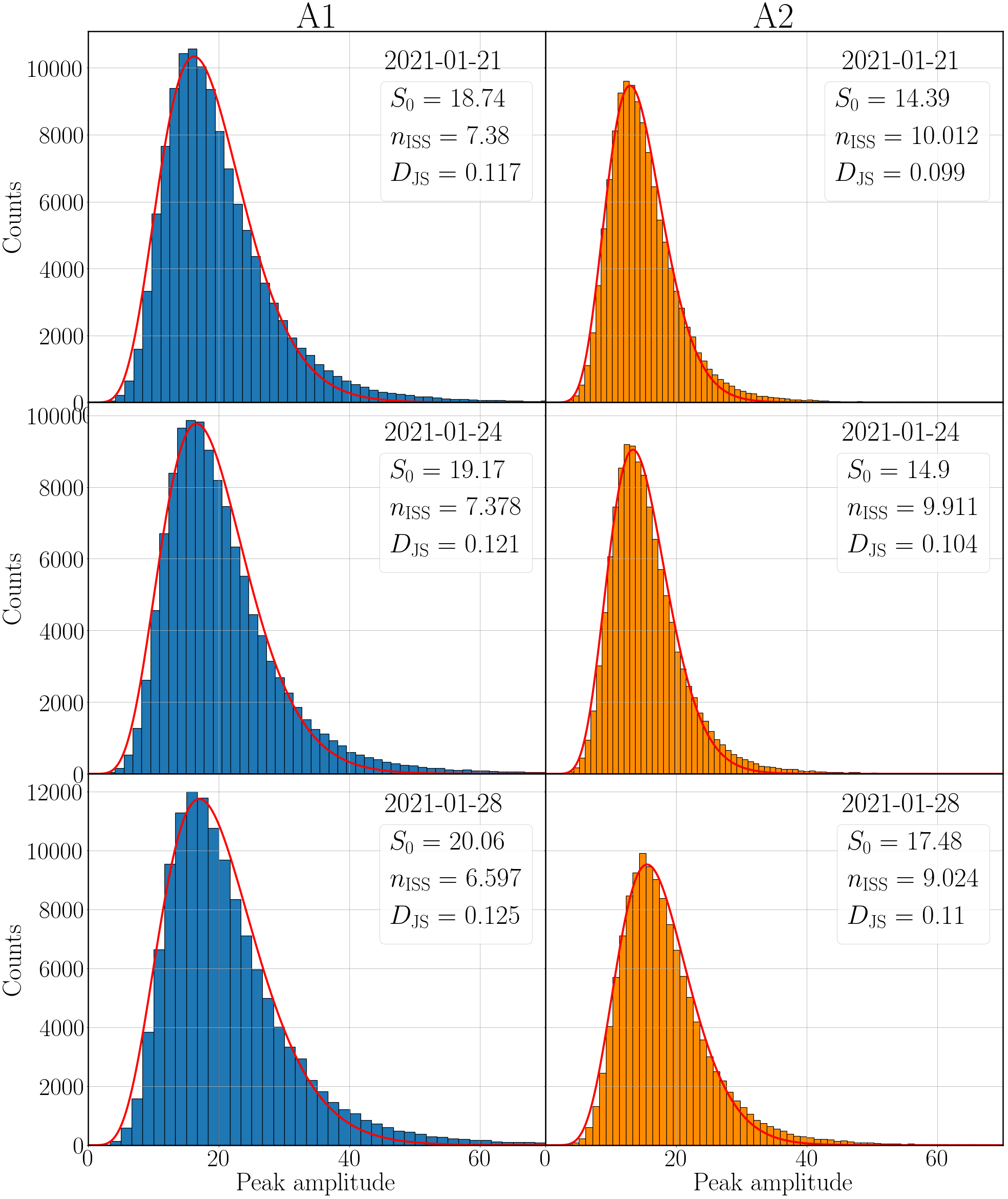}
    \caption{Histograms of projected pulse amplitude for J0835$-$4510 for A1 (left column) and A2 (right column) for the January 2021 observations. The curve shows the estimated scintillation distribution from fitting $n_\mathrm{ISS}$ in Eq.~(\ref{eq:scintillation}).}
    \label{fig:scintillations}
\end{figure}
Scintillation due to the interstellar media can change the intensity of the pulses.
Fig.~\ref{fig:scintillations} shows a histogram of the projected pulse S/N for A1 and A2. The line shows the estimated probability density function (PDF) from scintillation \citep{Cordes:1997my}, 
\begin{equation}
    f_S(S|n_\mathrm{ISS}) = \frac{\left(S\, n_\mathrm{ISS}/S_0 \right)^{n_\mathrm{ISS}}}{S\,\Gamma(n_\mathrm{ISS})}\exp{\left( \frac{-S\, n_\mathrm{ISS}}{S_0} \right)} \; \Theta(S),
    \label{eq:scintillation}
\end{equation} 
where $n_\mathrm{ISS}$ is the number of scintles, $S_0$ is the mean value of the signal $S$ (i.e., $S_0 = \langle \mathrm{S/N} \rangle$), and $\Theta$ is the Heaviside step function. Since $S \propto T_{\mathrm{peak}}$ \citep{lorimer2012handbook}, it follows that $T_{\mathrm{peak}}$ also obeys the PDF in Eq.~(\ref{eq:scintillation}).
Here we will explore the possibility of modeling the individual pulses amplitude distribution entirely in terms of a pure interstellar scintillation effect.
We therefore calculate $n_\mathrm{ISS}$ by fitting the observed single pulse peak amplitudes for each Radio telescope. We also normalize the number of observations in each bin by the total number of single pulses in each observation.

As a result, for Vela, we find $n_\mathrm{ISS} \sim 6.6 - 7.4$ with A1 and $n_\mathrm{ISS} \sim 9 - 10$ for A2. 
We also note the large value of the $n_\mathrm{ISS}$ found in comparison to the typical $n_\mathrm{ISS}<2$ found for longer time scales and different radio-frequencies. \citep{2000astro.ph..7233C} found two scintillation scales observing Vela at 2.5~GHz of 15s and 26s. Rescaling those scales to our observing frequency, 1400~MHz, we find time scales of $\Delta t_\mathrm{d,1} = 7.48$~s and $\Delta t_\mathrm{d,2} = 12.97$~s. Likewise we rescale the scintillation bandwidths
to 1400~MHz, and find $\Delta \nu_\mathrm{d,1} = 3.84$~MHz and $\Delta \nu_\mathrm{d,2} = 6.49$~MHz, respectively. We can now compare 
our values of $n_\mathrm{ISS}$ with theoretical estimations via the formula \citep{Cordes:1997my}
\begin{equation}
    n_\mathrm{ISS} \approx \left(1+\eta_t\frac{T}{\Delta t_\mathrm{d}}\right)\left(1+\eta_\nu\frac{BW}{\Delta \nu_\mathrm{d}}\right)
\end{equation}
where $\eta_t$ and $\eta_\nu$ are filling factors $\sim 0.25$. The estimated $n_\mathrm{ISS}$ for $T=0.089$~s A1 (BW~=~112~MHz) and for A2 (BW~=~56~MHz) are
$n_\mathrm{ISS,1}=8.3$ and $n_\mathrm{ISS,1}=4.7$ for A1 and A2 respectively, and $n_\mathrm{ISS,2}=5.3$ for A1 and $n_\mathrm{ISS,2}=3.2$ for A2. 

We then conclude that $n_\mathrm{ISS}$ over a shorter (0.089 seconds) timescale is expected to be smaller than measured for A2 and polarization dependent (values roughly match the one-polarization measures of A1). We also find a relatively good agreement between the observational data and the theoretical PDF, showing that Eq.~(\ref{eq:scintillation}) holds valid even at such short timescales. Nonetheless, we also note an excess in the number of high-amplitude single pulses that cannot be explained solely on the basis of scintillation. Those represent several thousands of pulses, and leave room for its interpretation in terms of pulsar intrinsic mini-giant pulses.

\subsubsection{Self-Organizing Map (SOM) techniques and results}\label{sec:CS}

\begin{figure*}[htb]
   \centering\includegraphics[width=\textwidth]{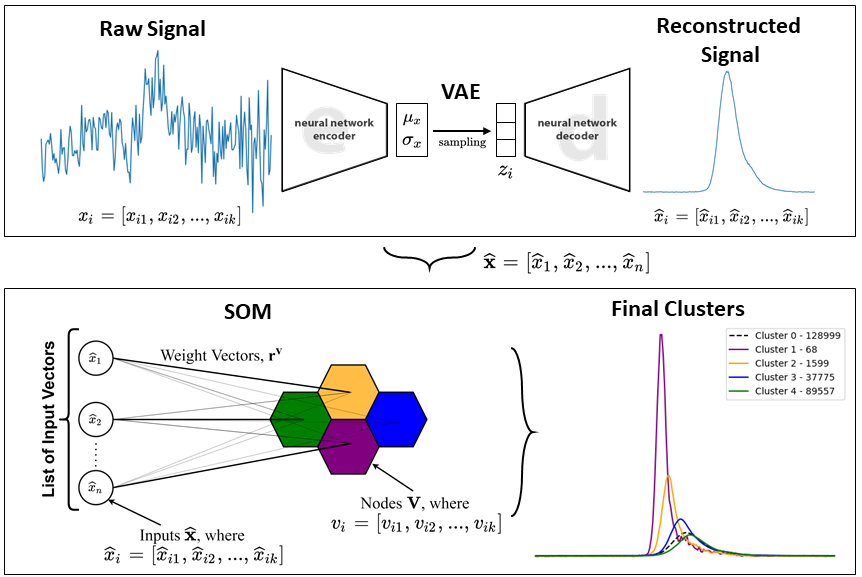}
   \caption{Schematics of the VAE reconstruction of all $N$ individual pulses signals and of SOM clustering.}
    \label{fig:VAESOM}
\end{figure*}

In deep learning literature, there are various unsupervised approaches to learn or capture the representations of the data. The most common ones include the autoencoder and its variants, a class of deep learning algorithms that take in the input and try to reconstruct the same input by passing it through the low-dimensional bottleneck subjected to different regularizations (e.g., sparsity). In our case, we consider a variational autoencoder (VAE) which is a probabilistic model with stochastic latent space \citep{kingma2014autoencoding}.

After training the VAE, we consider the Self-Organizing Map (SOM) for unsupervised clustering. SOM is a type of neural network that produces a low-dimensional map (2D), a discretized representation of the input samples. We present the schematic diagram of VAE and usage of SOM for clustering in Fig. \ref{fig:VAESOM}.

\begin{figure}[htb]\begin{center}
	\includegraphics[width=\columnwidth]{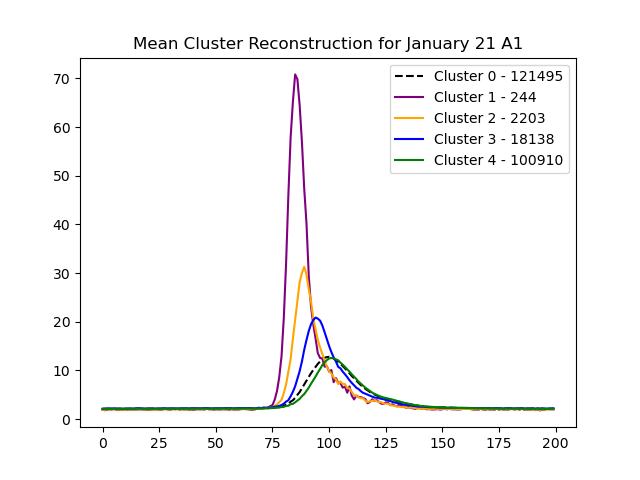}\vskip-10pt
	\includegraphics[width=\columnwidth]{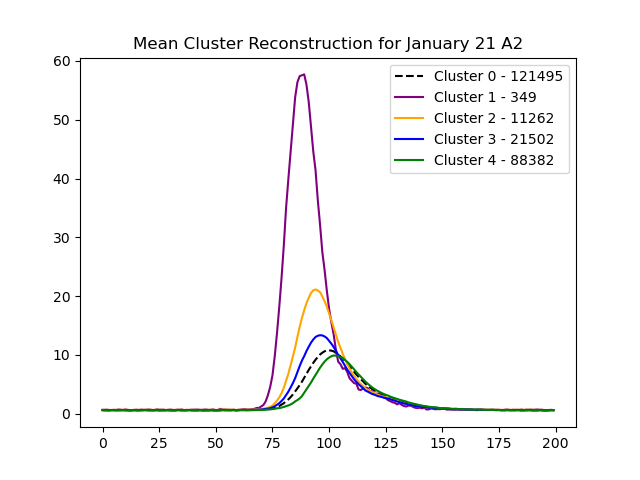}
    \caption{Distribution of the SOM clustering average signals for the observations 2021-01-21. Radio telescope A1 on top and Radio telescope A2 on bottom with respective VAE training performed on the January 28 observation.}
    \label{fig:1A12CS}
    \end{center}
\end{figure}

In Fig.~\ref{fig:1A12CS} we display, as a sample, the average value of the pulses in each SOM cluster for each Radio telescope observation on 2021-01-21. Those have been obtained by first applying a reconstruction of the raw pulses with the VAE technique, for which we have used the reconstruction of our best day of observation (2021-01-28) as a training case to apply to the rest of the days of observation. This training has been applied for each antenna individually. SOM allows us to specify then the number of clusters we seek to subdivide the whole set. We have studied several possible cases, 4, 6, 10, 25, 100 clusters, finding that the simplest four cluster analysis presents the most robust results. 

We use the SOM clustering to determine four relevant sets of pulses characteristics that seem to linearly array along the emission regions of the magnetosphere (See Fig.~\ref{fig:altitudes}) separated by roughly $100$km each. 

\subsubsection{Geometrical modeling of the cluster components}\label{sec:magneto}

\begin{figure}[htb]
	\includegraphics[width=\columnwidth]{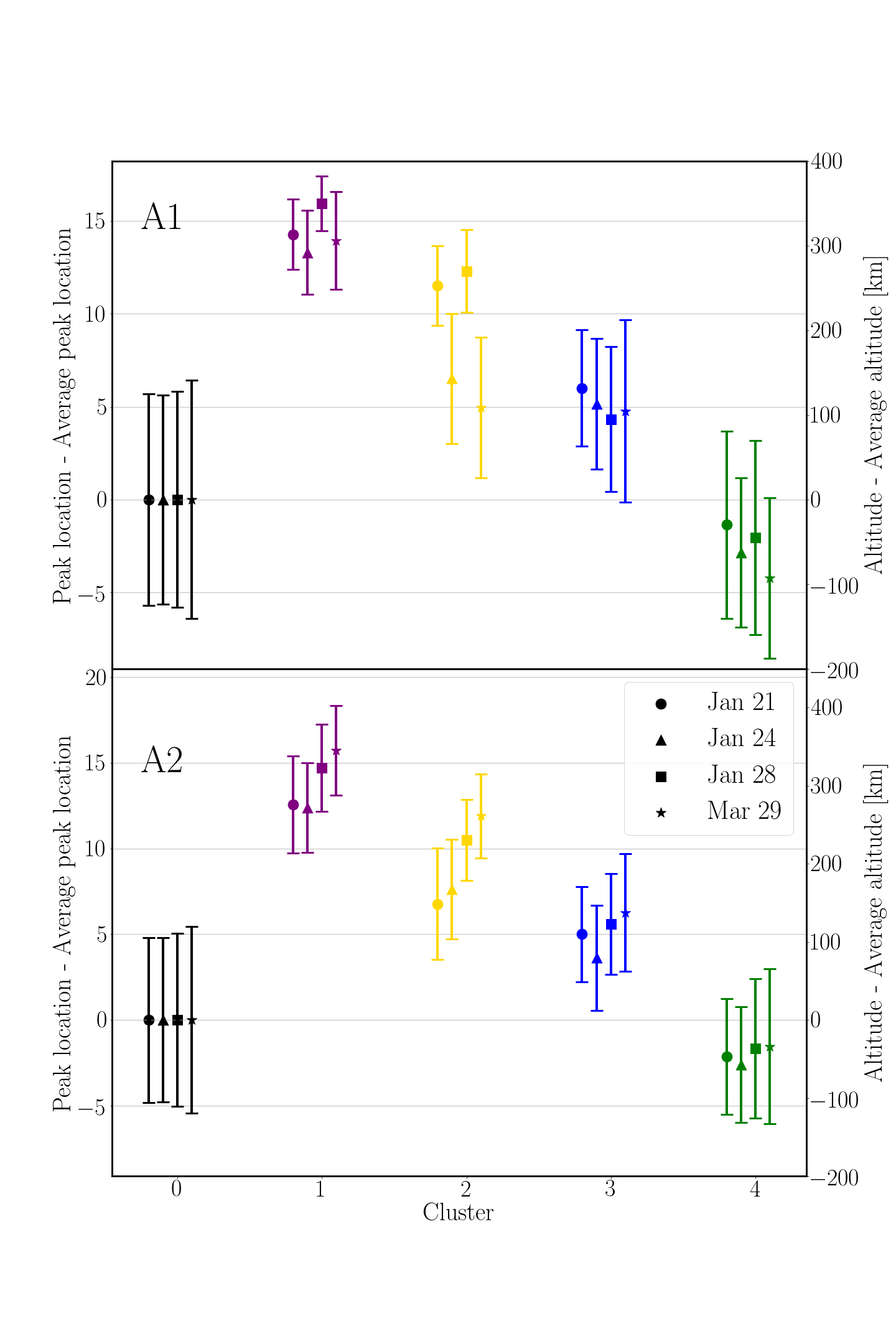}
    \caption{Peak location and magnetosphere altitude, with the corresponding error bars, for each of the pulse clusters presented in Sec.~\ref{sec:CS}.}
    \label{fig:altitudes}
\end{figure}

As presented Sec.~\ref{sec:scintillations}, the excess of high amplitude pulses cannot be explained solely due to effects of scintillation. In Sec.~\ref{sec:CS} we also find that each cluster has a different average peak location, with brighter pulses arriving earlier. Therefore, following the classic work of \citep{Downs1983},
we may attribute these variations in pulse amplitude and location to different altitudes in the neutron star magnetosphere where the pulses of each cluster are emitted. To this end, we measure the displacement of each cluster peak location relative to the average pulse location, and then relate those pulse displacements to differences in the emission altitude by
\begin{equation}
    h - \bar{h} = \frac{x - \bar{x}}{n_{bins}} c P ,
\end{equation}
where $x$ and $h$ are the cluster peak location and altitude in the magnetosphere, $\bar{x}$ and $\bar{h}$ are the average peak location and the average altitude (corresponding to cluster 0), $n_\mathrm{bins}$ is the number of time bins in each pulse (in our case, $1220$), and $P$ is the pulsar rotational period. 
Fig.\ref{fig:altitudes} displays the results of applying this model to each of the four days of observation for each antenna. The right hand side ordinate gives
the components distances to the average pulse reference height in the pulsar magnetosphere. We note the consistency between the components for each of the four days and for each individual antenna's observations. The four components appear to be almost equidistant (this maybe an effect of the SOM clustering method) and roughly of the order of $\sim100$ kilometers.

\subsection{Glitching pulsars monitoring program}\label{sec:glitches}

The PuMA collaboration has been monitoring with high cadence a set of pulsars from the southern hemisphere that had shown glitches before \citep{Gancio:2019frj}.
The Vela Pulsar (PSR B0833$-$45 / PSR J0835$-$4510) is one of the most active pulsars in terms of glitching, counting 21 in the last 50+ years.
Our early monitoring allowed us to detect a large glitch on 2019 February 1st \citep{atel_vela}, measured with observations three days before and three days after the event.
In addition to Vela, we are currently systematically monitoring the pulsars mentioned in \citep{Gancio:2019frj}, 
J0742$-$2822, J1048$-$5832, 
J1644$-$4559, 
J1721$-$3532, J1731$-$4744, J1740$-$3015, 
and plan to extend this list to other accessible (bright) glitching pulsars. 

In \citep{Zubieta:2022umm}  we reported on the new results of a systematic monitoring of southern glitching pulsars at the Argentine Institute of Radioastronomy that started in the year 2019. We detected a major glitch in the Vela pulsar (PSR J0835$-$4510) and two small-glitches in PSR J1048$-$5832. For each glitch, we presented the measurement of glitch parameters by fitting timing residuals. We then made an individual pulses study of Vela in observations before and after the glitch. We selected 6 days of observations around the major glitch on 2021 July 22 and study their statistical properties with machine learning techniques. We used Variational AutoEncoder (VAE) reconstruction of the pulses to separate them clearly from the noise. We performed a study with Self-Organizing Maps (SOM) clustering techniques to search for unusual behavior of the clusters during the days around the glitch not finding notable qualitative changes.
We have also detected and confirmed recent glitches in PSR J0742$-$2822 and PSR J1740$-$3015.

When a glitch occurs, the pulsar suffers a sudden jump in its rotation frequency. This spin up can be introduced in the timing model as a change in the phase of the pulsar modeled as \citep{glitch-timing}
\begin{eqnarray}\label{eq:glitch-model}
    \phi_\mathrm{g}(t) = \Delta \phi + \Delta \nu_\mathrm{p} (t-t_\mathrm{g}) + \frac{1}{2} \Delta \dot{\nu}_\mathrm{p} (t-t_\mathrm{g})^2 + \nonumber\\  \frac{1}{6} \Delta \ddot{\nu}(t-t_\mathrm{g})^3+
    \left[1-\exp{\left(-\frac{t-t_\mathrm{g}}{\tau_\mathrm{d}}\right)} \right]\Delta \nu_\mathrm{d} \, \tau_\mathrm{d},
\end{eqnarray}
where $\Delta \phi$ is the offset in pulsar phase, $t_\mathrm{g}$ is the glitch epoch, and $\Delta \nu_\mathrm{p}$, $\Delta \dot{\nu}_\mathrm{p}$ and $\Delta \ddot{\nu}$ are the respective permanents jumps in $\nu$, $\dot\nu$ and $\ddot{\nu}$ relative to the pre-glitch solution. Finally, $\Delta \nu_\mathrm{d}$ is the transient increment in the frequency that decays on a timescale $\tau_\mathrm{d}$. From these parameters one can calculate the degree of recovery, $Q$, which relates the transient and permanent jumps in frequency as $Q=\Delta \nu_\mathrm{d} / \Delta \nu_\mathrm{g}$. At last, two commonly used parameters in the literature are the instantaneous changes in the pulse frequency and its first derivative
(at the glitch epoch), which can be described as 
\begin{eqnarray}
    \Delta \nu_\mathrm{g} &= \Delta \nu_\mathrm{p} + \Delta \nu_\mathrm{d} \\ 
    \Delta \dot{\nu}_\mathrm{g} &= \Delta \dot{\nu}_\mathrm{p} - \frac{\Delta \nu_\mathrm{d}}{\tau_\mathrm{d}} \, .
\end{eqnarray}

\begin{figure}[htb]
    \centering
    \includegraphics[width=\linewidth]{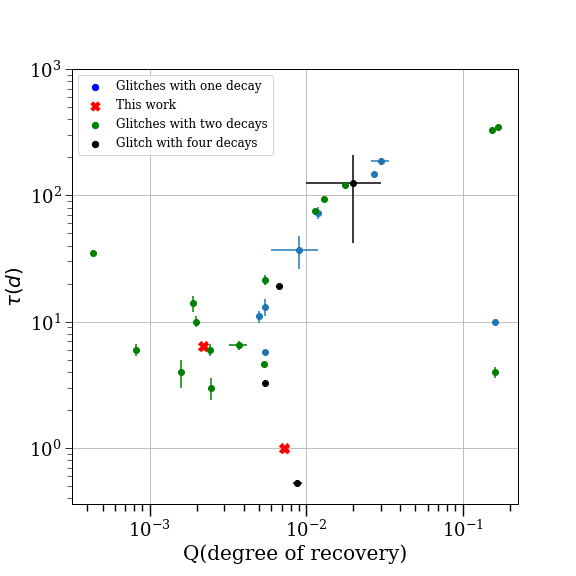}
    \caption{Comparison of current and previous glitches decaying parameters for Vela pulsar.}
    \label{fig:Qtau}
\end{figure}

Here we briefly review the detailed analysis \citep{Zubieta:2022umm} of the latest $(\#22$ recorded) 2021 large Vela glitch, with $(\Delta\nu_\mathrm{g}/\nu)_{2021}=1.2\times10^{-6}$,
providing an accurate description of the glitch characteristic epoch, jumps, and exponential recovery of 6.4 and 1 days times scales,
(See Table~\ref{tab:Vglitch} and Fig.~\ref{fig:Qtau}).

\begin{table}
  \centering    
  \caption{Timing model for the 2021 July 22nd Vela glitch}
   \begin{tabular}{ll}        
     \hline
     Parameter & Value \\
     \hline
     $\mathrm{PEPOCH}$ (MJD)&59417.6193 \\
     $\mathrm{F0}(\mathrm{s^{-1}})$& 11.18420841(1)\\
     $\mathrm{F1}(\mathrm{s^{-2}})$ &  $-1.55645(4)\times 10^{-11}$\\
     $\mathrm{F2}(\mathrm{s^{-3}})$ &  $6.48(1)\times 10^{-22}$\\
     $\mathrm{DM}(\mathrm{cm^{-3} pc})$ & 67.93(1)\\
     $t_\mathrm{g}$ (MJD) &59417.6194(2)\\
     $\Delta\nu_\mathrm{p}$ (s$^{-1}$) &1.381518(1) $\times 10^{-5}$\\
     $\Delta\dot{\nu}_\mathrm{p}$ (s$^{-2}$)&$-8.59(4)\times 10^{-14}$\\
     $\Delta\ddot{\nu}$\quad (s$^{-3}$)&$1.16(3)\times 10^{-21}$\\
     $\Delta\nu_\mathrm{d1}$\, (s$^{-1}$)&$3.15(12)\times 10^{-8}$\\
     $\tau_\mathrm{d1}$ (days) & 6.400(2)\\
     $\Delta\nu_\mathrm{d2}$\, (s$^{-1}$)&$9.9(6)\times 10^{-8}$\\
     $\tau_\mathrm{d2}$ (days) & 0.994(8)\\
     $\Delta \phi$ & $\sim0$\\
     $\Delta\nu_\mathrm{g}/\nu$ & $1.2469(5)\times 10^{-6}$\\
     $\Delta\dot\nu_\mathrm{g}/\dot\nu$ & 0.084(5)\\
     $Q_1$ & 0.00226(9)\\
     $Q_2$ & 0.0071(4)\\
     \hline
   \end{tabular}
  \label{tab:Vglitch}
 \end{table}

\subsubsection{Machine Learning analysis of the Vela Glitch Day:  2021, July 22 observations with A2}\label{sec:July22}

The observations on 2021 July 22 (the day of the glitch) with A2 are divided in three data parts. The first of those observations, starting at MJD 59417.65584, is about 52 minutes after the estimated occurrence of the glitch at MJD 59417.6194(2). The total observation time on July 22 is 2.65~h (divided into three observations) with a total SNR of 689.
Since those three individual sub-observations contain enough pulses each to make a SOM analysis, we proceed to consider them individually and independently. 
In search for more subtle details, we choose a six clusters study.

\begin{figure*}[htb]
   \centering 
\includegraphics[width=\textwidth]{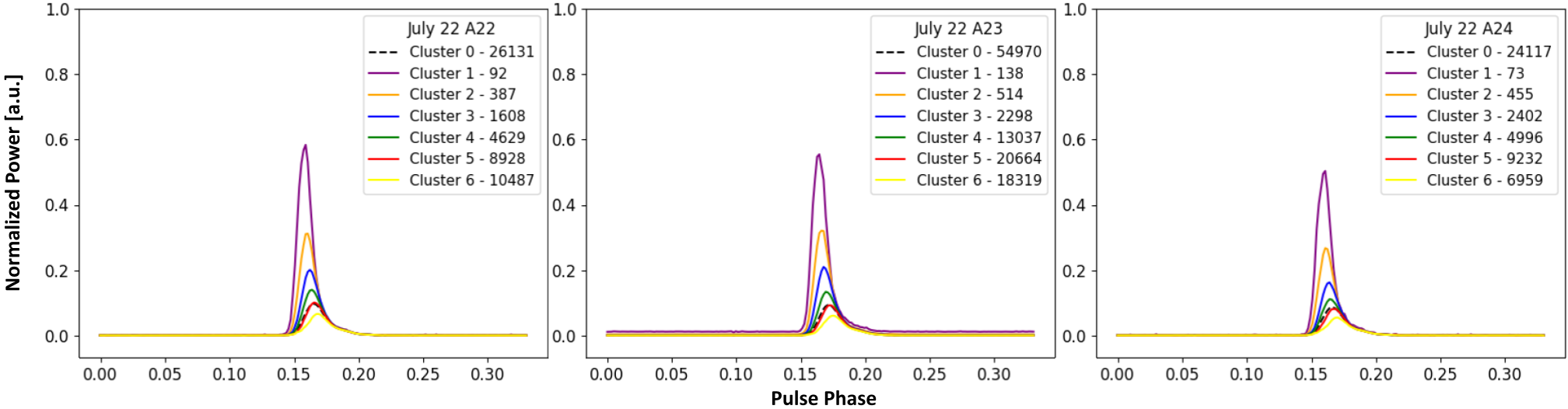}
     \caption{Mean cluster pulses for  2021 July 22 three successive observations (roughly from 1 to 3.5~h after the glitch) with Antenna 2 for 6 SOM clusters with VAE reconstruction. 200 (out of total 611) phase bins were taken around the mean peak of each day to perform the single-pulse analysis on.}
  \label{fig:7-22A2}
 \end{figure*}

The results of those 6 SOM clustering studies are displayed in Fig.~\ref{fig:7-22A2}. 
We first observe that the right wing side of each mean cluster pulse seem all to superpose and that the sequence of those mean pulse clusters, with increasing amplitude, seem to appear earlier and earlier on average. The pulse width also shows a (weak) dependence on the cluster, being narrower for higher amplitude mean pulses. All these features, for the three observations covering from roughly 1--3.5~h  after this large glitch seem to be similar to those well in between glitches, as we have observed in our previous analysis of the Vela pulses from January and March 2021 \citep{Lousto:2021dia}.

\subsubsection{Other glitching pulsars}\label{sec:OGP}

The accuracy of our observations and procedures
allowed us to determine two mini-glitches (the smallest recorded so far) in PSR 1048$-$5832, $(\#8$ and $\#9$ recorded), with $(\Delta\nu_\mathrm{g}/\nu)_{2020}=8.9\times10^{-9}$ and $(\Delta\nu_\mathrm{g}/\nu)_{2021}=9.9\times10^{-9}$, respectively.


On 2022 September 21, MJD=59839.4(5), a new glitch \#9 in PSR J0742$-$2822 was reported by
\citep{2022ATel15622....1S}. We have been able to confirm this glitch with our data \citep{2022ATel15638....1Z} and find relative jumps of $\Delta \nu / \nu = 4.29497(2) \times 10^{-6}$ and $\Delta \dot\nu / \dot\nu = 0.0510(7)$, making it the largest recorded glitch for this pulsar.

Also, on 2022 December 22, MJD=59935.1(4), we detected a new glitch in PSR J1740$-$3015 that was reported in \citep{2022ATel15838....1Z}.
We found a relative jump of $\Delta \nu / \nu = 3.32(3) \times 10^{-7}$ and plan 
to continue monitoring PSR J1740$-$3015 to improve the post-glitch timing solution. 

(See E. Zubieta's contribution to this volume for other details on the glitching pulsar observations mentioned here).


This concludes our review of the main PuMA results and we now discuss the new projects underway.


\section{Future Projects and Conclusions}\label{sec:Conclusions}


\subsection{Single-pulse timing}\label{sec:SPT}

We presented the first detailed analysis of the 2019/2020 observational campaign towards the bright millisecond J0437$-$4715 using the two antennas at IAR's observatory. This data set comprises now over three additional years of high-cadence (up to daily) observations with both antennas, A1 and A2. By both improving the time baseline of observations, (including the additional three years of data) and combining it with a future single pulse analysis, we may achieve up to factors 2-3 improvements in the timing accuracy \citep{Kerr:2015dla}. 
We recall here that one of the main results of the Vela single pulse analysis of \citep{Lousto:2021dia} was the narrower nature of the high amplitude pulses clusters, which, if translated to J0437$-$4715, would lead to notable timing improvements.

It is worth noting here that there are several difficulties to overcome in order to further dramatically improve the timing of millisecond pulsars at lower frequencies, among them the accurate modeling of the dispersion and Faraday's rotation of the interestellar media scaling like the inverse frequency squared, $1/\nu^{2}$, and the scattering and scintillation scaling like $1/\nu^{4.4}$. Additionally one has to account for the time dependence of the Dispersion measure (DM), solar wind, and the frequency evolution of the pulsar profile.

Ongoing and future hardware upgrade of IAR's antennas, such as installing larger-bandwidth boards (from the current 56MHz (Ettus\footnote{\url{https://www.ettus.com/all-products/usrp-b205mini-i-board/}}) to 400MHz (ROACH\footnote{\url{https://www.digicom.org/roach-board.html}}) ), promise to expand IAR's observational capabilities and improve its achievable timing precision by raising sensitivity at least by a factor 3. Thus,
with the future improvements in IAR’s antennas receivers, which include a combination of broader bandwidth and reduction of system temperature, it will be possible to study the dynamical spectra of single pulses for other pulsars of interest, such as the glitching PRS J1644-4559 and J0738-4042, and the millisecond pulsar J0437-4715 to contribute to improve pulsar timing arrays data in order to detect a stochastic gravitational waves background. We display some to the pulse properties of this primary choice of pulsars accessible to IAR for single-pulse studies in Table~\ref{tab:SPT}.

\begin{table*}
  \centering    
  \caption{IAR's accessible pulsars for single-pulse studies.}
   \begin{tabular}{llllllllll}        
     \hline
PSR  & P0 & S1400 & G & W50 & W10 & P0/W50 & P0/W10 & S/N & S/N \\
J2000 & (s) & (mJy) & $\#$  & (ms) & (ms) & & &(W50) &(W10) \\
     \hline
J0437-4715  &  0.005757  & 150.20 & * & 0.141  & 1.020 & 40.83299 & 5.644561 & 947.963 & 323.6997 \\
J0738-4042 & 0.374921 & 112.60 & * & 25.000 & 39.000 & 14.99683 & 9.613352  & 421.2629 & 330.4643 \\
J0835-4510 & 0.089328 & 1050.00  & 22 &  1.700 & 3.800 & 52.54611 & 23.50747 & 7538.54 & 4981.414 \\
J1644-4559 & 0.455078 & 300.00   &  4 & 8.000  &  13.323  & 56.88478 & 34.15734 & 2242.684 & 1727.472 \\
     \hline
   \end{tabular}
  \label{tab:SPT}
 \end{table*}

We recall here the standard formula for the expected signal-to-noise ratio \citep[S/N;][]{lorimer2012handbook}:
\begin{equation} \label{eq:snr}
    \mathrm{S/N} \lesssim S_\mathrm{mean}  \frac{\sqrt{n_\mathrm{p}  t_\mathrm{obs} B}}{G T_\mathrm{sys}} \sqrt{\frac{P-W}{W}},
\end{equation}
where $S_\mathrm{mean}$, $P$, and $W$ are the mean flux density, period and equivalent width of the pulses, respectively, $G$, $B$, $n_\mathrm{p}$, and $T_\mathrm{sys}$ are the antenna gain, bandwidth, number of polarizations, and system temperature, respectively; with $t_\mathrm{obs}$, the effective observing time. 
In table~\ref{tab:SPT} we used the data from the ATNF catalogue (\url{http://www.atnf.csiro.au/people/pulsar/psrcat/}) to estimate a relative $S/N=S1400\sqrt{P/W-1}$ (arbitrary normalization) and $P0/W$ as a measure of the relative extension of the pulse over the period. We use the notation 
PSRJ: Pulsar name based on J2000 coordinates, 
P0: Barycentric period of the pulsar (s), 
S1400:       Mean flux density at 1400 MHz (mJy),
G (NGlt):    Number of glitches observed for the pulsar,
W50:         Width of pulse at 50\% of peak (ms),
W10:         Width of pulse at 10\% (ms).

\subsection{Intelligent Fast Radio Burst seaches}\label{sec:FRB}

With thousands of good quality hours of pulsar observations it is interesting to see if the data contains also FRB \citep{2019A&ARv..27....4P} signals. In particular, machine learning techniques have been developed to perform massive searches of FRB \citep{Zhang:2018jux} and its classification \citep{Connor:2018wfr,Wagstaff:2016pdw}
using supervised \citep{Luo:2022smj} and unsupervised methods \citep{Chen:2021jpq,Zhu-Ge:2022nkz}. 
A most practical implementation for fast transient classification\citep{2020MNRAS.497.1661A} is the Fetch code: \url{https://github.com/devanshkv/fetch}.
Another useful tool is the synthetic FRB generator: \url{https://gitlab.com/houben.ljm/frb-faker} to train FRB searches and classification.
A Living Theory Catalogue for Fast Radio Bursts with a review of the numerous existing theories to model FRBs is reported in \citep{Platts:2018hiy}.

(For more on magnetars and FRB projects at IAR, see also S.B. Araujo Furlan's presentation in this volume).

\subsection{Lower frequencies observations}\label{sec:MIA}

IAR's Multipurpose Interferometer Array (MIA) is the only project of its kind in South America: a versatile low frequency interferometer (100 MHz - 2 GHz) designed to investigate transient sources and non-thermal cosmic radiation from the Southern Hemisphere. 
MIA will initially consists of a 16 antennas array of 5 meters in diameter each, distributed over a baseline of 50Km in order to obtain an angular resolution of at least 1.5 seconds of arc in the L-Band.

\begin{figure}[htb]\begin{center}\vskip 10pt
	\includegraphics[width=0.65\columnwidth]{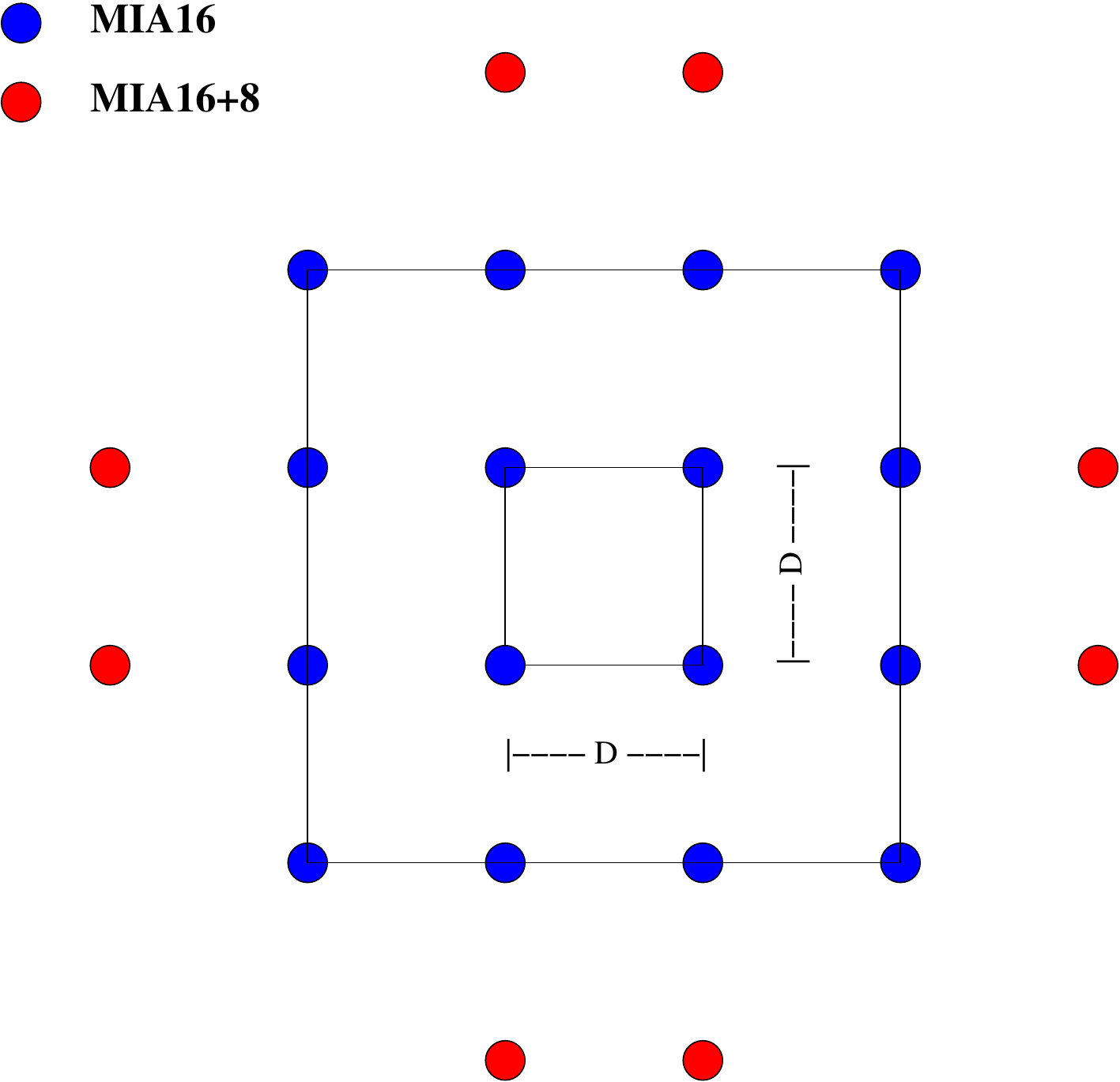}\vskip 10pt
	\includegraphics[width=0.55\columnwidth]{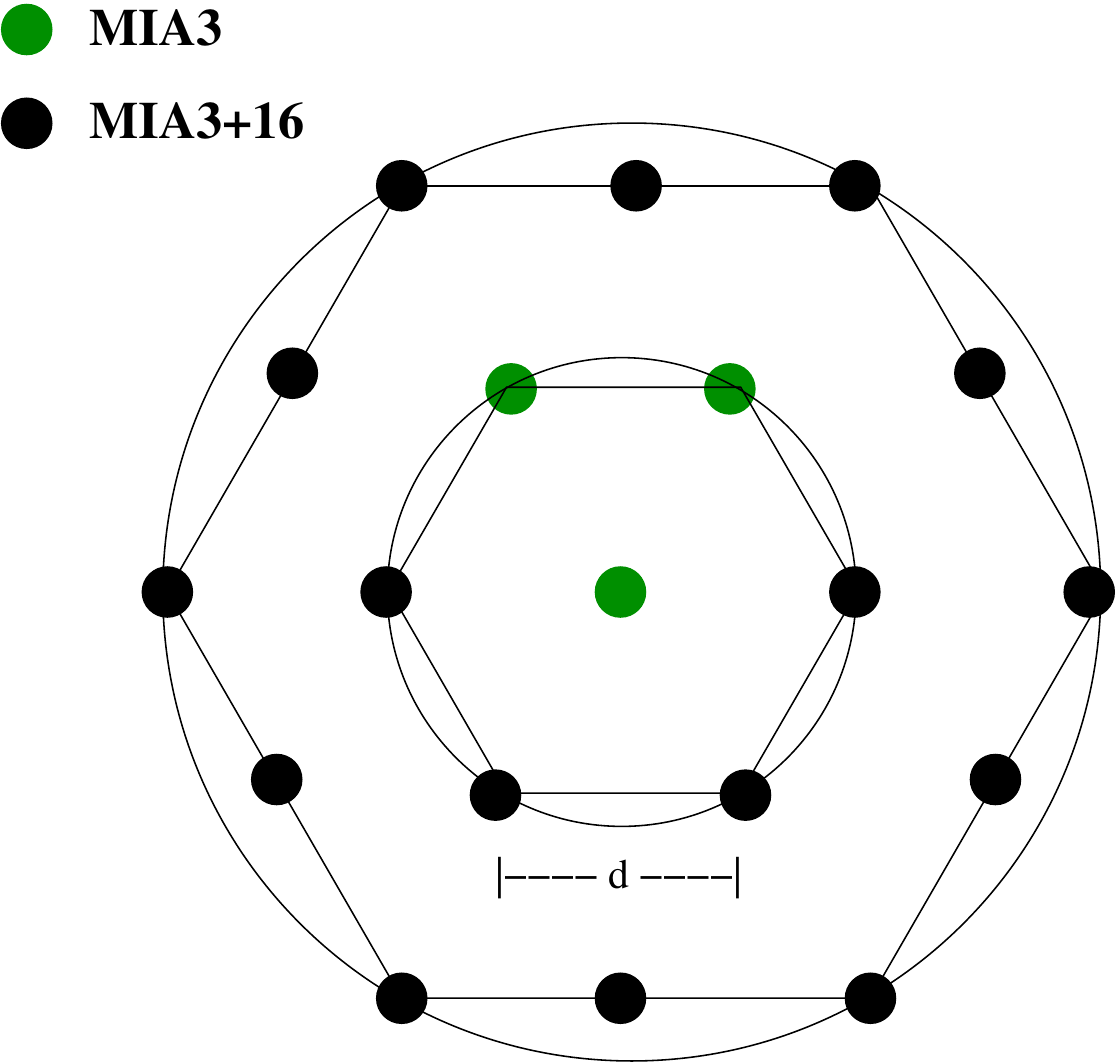}
    \caption{On top the MIA24 configuration and on bottom an alternative MIA19 for interferometric pulsar observations.}
    \label{fig:MIA}
    \end{center}
\end{figure}

Regarding the use of MIA for pulsar observations, we may consider that
in order for MIA to have a collecting surface equivalent to the 30 meters diameter A1 and A2 at IAR, one should have 36$\times$5 meters MIA antennas. This number can be reduced to about 24 dishes if they use solid slabs instead of wired ones, since the former raise the gain to about 60\% from the later 40\%. Another gain can be obtained if individually optimized dedicated receivers can be used for each  frequency sub-bands 
0.3-1 GHz\footnote{https://www.pasternack.com/5-section-high-pass-filter-300-mhz-1000-mhz-passband-700-mhz-pe8718-p.aspx},
1-2 GHz\footnote{https://www.pasternack.com/11-section-band-pass-filter-1-2-ghz-passband-1000-mhz-pe87fl1012-p.aspx}, and
2-4 GHz\footnote{https://www.pasternack.com/11-section-band-pass-filter-2-4-ghz-passband-2-ghz-pe87fl1013-p.aspx}.
Those 24 MIA antennas could be developed in three stages of 4 pre-series + 12 series + 8 additional to complete an hexagonal pattern of two 3 by 4 sub-arrays as displayed on top of Fig.~\ref{fig:MIA}. For instance, setting those 24 MIA antennas $D=2$km apart would cover about a land region of $S=15D^2=60$km$^2$ and a baseline of $B=5D=10$km.
This square blocks configurations of 4 antennas allow for an interferometric locking with total cancellation of phase and amplitude. 
An alternative configuration, based on the initial prototype of 3 antennas, is to place them in equilateral triangles in an hexagonal pattern as displayed in the bottom Fig.~\ref{fig:MIA}. The MIA19 with separations of $d=2$km between dishes would cover a ground area of about $S=6\sqrt{3}d^2=42$km$^2$ and a radial baseline of $B=4d=8$km. 
The basic interferometric pattern would thus be similar to that studied for the space antenna LISA with three equilaterally placed spacecrafts in orbit around the sun \citep{LISA:2017pwj}. 

By accessing lower frequencies, MIA would allow to include new glitching pulsars to our survey. Pulsars that typically have negative spectral index emit much more in the 400MHz band than in the 1400MHz band we use to observe with A1 and A2.
Using the ATNF data base we can find (See Table \ref{tab:MIA}) two dozen glitching pulsars with sufficient S/N to be observed by MIA that currently cannot be acurately followed with A1 and A2. The angular resolution of MIA in the configurations of Fig.~\ref{fig:MIA} with $B>1$km, would beat the angular resolution of 1' versus the current 30' of A1 and A2 at 1400MHz.
\begin{table}
  \centering    
  \caption{MIA's accessible glitching pulsars}
   \begin{tabular}{llllll}        
     \hline
PSR&P0&S400&S1400&G&S/N\\
J2000&(s)&(mJy)&(mJy)&$\#$&@400\\
     \hline
     J0820-1350  &   1.238  &  102.0   &   6.00     &    2  &        741.4 \\
     J1602-5100  &   0.864  &   45.0   &   8.23   &      1  &        643.9 \\
     J1836-1008  &   0.563  &   54.0   &   4.80   &      1   &       458.4 \\
     J1740-3015  &   0.607  &   24.6   &   8.90    &    37    &      349.0 \\
     J1703-4851  &   1.396   &  22.0   &   1.40     &    1    &      246.9 \\
     J1835-1106  &   0.166  &   30.0   &   2.50    &     1    &      193.4 \\
     J1705-1906  &   0.299  &   29.0   &   5.66    &     4    &      172.7 \\
     J1720-1633  &   1.566  &   13.0   &   1.10    &     1    &      147.9 \\
     J1705-3423  &   0.255  &   31.0   &   5.30    &     3    &      139.6 \\
    J1328-4357   &  0.533  &   18.0   &   4.40    &     1     &     123.9 \\
    J0846-3533  &   1.116  &   16.0   &   5.00    &     1     &     121.6 \\
    J1257-1027  &   0.617  &   12.0   &   1.20    &     1    &      115.4 \\
    J1320-5359  &   0.279  &   18.0   &   2.10    &     2    &      105.6 \\
    J1757-2421  &   0.234  &   20.0   &   7.20    &     1    &      94.68 \\
    J0758-1528   &  0.682  &    8.2   &   2.60    &     1     &     78.85 \\
    J0905-5127   &  0.346  &   12.0  &    1.05   &      2     &     77.54 \\
    J1123-6259  &  0.271   &  11.0   &   0.51    &     1      &    70.79 \\
    J1803-2137   &  0.134  &   23.0   &   9.60    &     5     &     70.07 \\
    J0729-1836   &  0.510  &   11.2   &   1.90   &      2    &      66.67 \\
    J1824-2452A  &  0.003  &   40.0   &   2.30   &      1    &      58.54 \\
    J1141-3322   &  0.291  &    8.0   &   1.60    &     1    &      58.23 \\
    J1743-3150   &  2.415   &   6.6   &   2.10    &     1     &     47.89 \\
    J1801-2451   &  0.125   &   7.8   &   1.46    &     7     &     39.02 \\
    J1730-3350   &  0.139   &   9.2   &   4.30    &     3     &     36.81 \\
     \hline
   \end{tabular}
  \label{tab:MIA}
 \end{table}

(See G. Gancio's contribution to this volume for a description of the MIA project carried out at IAR).

\bigskip

Let us conclude here with a quote from a most inspiring author,
{\it Pensar es olvidar diferencias, es generalizar, abstraer.}
"To think is to forget a difference, to generalize, to abstract.”, 
from 'Funes the Memorious', Ficciones ― Jorge Luis Borges \citep{Borges}.

\section*{Acknowledgements}

COL gratefully acknowledge the National Science Foundation (NSF) for financial support from Grants No.\ PHY-1912632, PHY-2207920  
and RIT-COS 2021-DRIG grant. JAC and FG are CONICET researchers. JAC is a Mar\'ia Zambrano researcher fellow funded by the European Union  -NextGenerationEU- (UJAR02MZ). This work received financial support from PICT-2017-2865 (ANPCyT) and PIP 0113 (CONICET). JAC and FG were also supported by grant PID2019-105510GB-C32/AEI/10.13039/501100011033 from the Agencia Estatal de Investigaci\'on of the Spanish Ministerio de Ciencia, Innovaci\'on y Universidades, and by Consejer\'{\i}a de Econom\'{\i}a, Innovaci\'on, Ciencia y Empleo of Junta de Andaluc\'{\i}a as research group FQM-322, as well as FEDER funds.

\bibliography{refs}
\end{document}